\documentclass[prl,amsfonts,amssymb,floats,twocolumn,superscriptaddress,aps]{revtex4}

\usepackage{amsmath}
\usepackage{bm}
\usepackage{graphicx}

\def\H{{\cal H}}
\def\e{\epsilon}

\begin{document}

\title{Zero-bias conductance in carbon nanotube quantum dots}

\author{Frithjof B. Anders}
\affiliation{Institut f\"ur Theoretische Physik, Universit\"at Bremen,
                  P.O. Box 330 440, D-28334 Bremen, Germany}

\author{David E. Logan}
\author{Martin R. Galpin}
\affiliation{Physical and Theoretical Chemistry, Oxford
  University, South Parks Road, Oxford OX1 3QZ, UK}

\author{Gleb Finkelstein}
\affiliation{Department of Physics, Duke University, Durham, North
  Carolina 27708, USA}

\date{January 21, 2008}

\begin{abstract}
We present numerical renormalization group calculations for the
zero-bias conductance of quantum dots made from semiconducting carbon
nanotubes. These explain and reproduce the thermal evolution of the conductance for different groups of orbitals, as the dot-lead tunnel coupling is varied and the system evolves from correlated Kondo behavior to more weakly correlated regimes. For integer fillings $N=1,2,3$ of an  $SU(4)$ model, we find universal scaling behavior of the conductance that is distinct from 
the standard $SU(2)$ universal conductance, and concurs quantitatively with experiment. Our 
results also agree qualitatively with experimental differential conductance maps.

\end{abstract}

\maketitle

\paragraph*{Introduction:}

Carbon nanotubes have generated immense interest due to their rich
transport properties~\cite{Dekker1999}. Their small capacitance
generates a Coulomb blockade regime at low temperatures~\cite{Dekker1997}, and 
a single electron transistor~\cite{KastnerSET1992,Dekker1997} can be made out 
of weakly coupled nanotubes.  The Kondo effect, studied in transition metal
ions and rare earth compounds for some fifty years~\cite{Hewson93,Wilson75}, 
is now equally a classic hallmark of many-body physics in nanoscale devices, 
including carbon nanotubes~\cite{Nygard2000} as well as semiconducting quantum 
dots, molecules and magnetic adatoms on metallic surfaces (for a review 
see~\cite{Glazman2001}).

While the `standard' Kondo effect arising in nano-devices is the orbitally 
non-degenerate, spin-$\frac{1}{2}$ ($SU(2)$) Kondo effect~\cite{Hewson93}, 
it is known that in quantum dots made from high quality nanotubes, a series of 
\emph{two} spin-degenerate orbitals originates from the two electronic 
sub-bands~\cite{Liang2002,Buitelaar2002}. Consecutive filling of
these orbitals should thus yield an $SU(4)$-type Kondo 
effect~\cite{Hewson93,Jarillo2005,FinkelsteinI2007,FinkelsteinII2007,Bickers87,Boese2002,Borda2003,Galpin2005,Lopez2005,Choi2005,Mitchell2006,Busser2007}. 
This has indeed been observed in recent 
experiments~\cite{Jarillo2005,FinkelsteinI2007,FinkelsteinII2007},
including in particular systematic exploration of the $SU(4)$ conductance 
regimes~\cite{FinkelsteinII2007} via careful control of sample contact transparency.

In this paper, we present a consistent picture which explains the experimental zero bias conductance~\cite{FinkelsteinII2007}. We study the thermal evolution of the conductance as a function of applied gate voltage, for an $SU(4)$ Anderson model, using 
Wilson's numerical renormalization group (NRG)~\cite{Wilson75,BullaCostiPruschke2007}
(previous NRG work~\cite{Boese2002,Borda2003,Galpin2005,Lopez2005,Choi2005,Mitchell2006} 
has been confined to $T=0$); and show that on progressive increase of 
the dot-lead tunnel couplings, the system exhibits a rich evolution from correlated Kondo behavior with strong associated Coulomb blockade peaks in the zero-bias conductance, 
through to a weaker coupling regime where charge fluctuations are significant and Coulomb 
blockade oscillations suppressed. We obtain universal scaling functions for the zero-bias 
conductance, which deviate significantly from standard $SU(2)$ Kondo scaling (reflecting
the different universality class of our model); and show that this captures
experiment quantitatively.

\paragraph*{Theory:}

Interacting quantum dots, molecular junctions and other nano-devices
are modelled by the interacting region $\H_{imp}$, a set of
non-interacting reservoirs $\H_B$ and a coupling between the
subsystems $\H_T$: $\H = \H_{imp} + \H_{B} + \H_{T}$. For the carbon
nanotube quantum dot we restrict ourselves to the filling of  one
shell comprising two degenerate orbitals, and consider the local
Hamiltonian $\H_{imp} = \frac{1}{2}U(\hat{N}-N_{g})^{2}$. Here
$\hat N= \sum_{\alpha\sigma}d^{\dagger}_{\alpha\sigma}d^{\phantom\dagger}_{\alpha\sigma} = \sum_{\alpha\sigma}\hat{n}_{\alpha\sigma}$, where
$d^\dagger_{\sigma\alpha}$ creates a $\sigma$-spin electron
in orbital $\alpha =1,2$, the charging energy  $U=e^{2}/C \equiv E_{C}$ with 
dot capacitance $C$, and  $N_{g}$ denotes the dimensionless external gate voltage. 
Equivalently,
\begin{eqnarray}
   \H_{imp} &=& E_{d}\hat{N} + \frac{1}{2}U\sum_{m,m^{\prime}(m^{\prime} \neq m)}
    \hat{n}_{m}\hat{n}_{m^{\prime}}
\end{eqnarray}
with level energy $E_{d} = \frac{U}{2}(1-2N_{g})$ and $m \equiv (\alpha ,\sigma)$ a 
flavor index. For the isolated dot, with integer ground state charge
($\langle \hat{N} \rangle =$)
 $N_{d} = 0-4$, the 
edges in the Coulomb blockade staircase between charge $N_{d}$ and $N_{d}+1$ occur for 
half-integral $N_{g} = \frac{1}{2}(2N_{d}+1)$ (\emph{ie} $E_{d} = -UN_{d}$).
In Eq.(1) we have taken $U^{\prime}=U$, with $U$ [$U^{\prime}$] the intra- [inter-] orbital 
Coulomb repulsion, and have 
also 
neglected an exchange splitting, $J$. This is consistent 
with the argument~\cite{Oreg2000} that for carbon nanotubes, $U' \simeq U$ and
$J$ is much smaller than the charge fluctuation scale; and is known to be compatible with the 
nanotube data~\cite{Finkelstein2006}.

The tunneling to the two leads $\nu=L,R$  is assumed to be spin and orbital 
conserving~\cite{Choi2005,FinkelsteinII2007},
\begin{eqnarray}
   \H_{T} &=& \sum_{\alpha ,\nu} \tilde t_{\nu} \sum_{k,\sigma}
   \left(c^\dagger_{k\sigma\alpha\nu}d_{\sigma\alpha}^{\phantom\dagger}
 + d^\dagger_{\sigma\alpha}c_{k\sigma\alpha\nu}^{\phantom\dagger}
\right)
\end{eqnarray}
such that the overall $\H$ has $SU(4)$ symmetry. Only the binding combination of
lead states, 
$ c_{k\sigma\alpha} = \frac{1}{t}\sum_{\nu}\tilde{t}_{\nu}c_{k\sigma\alpha\nu}$ 
with $t = [\tilde{t}^2_{L}+\tilde{t}^2_{R}]^{\frac{1}{2}}$,
then couples to the dot. We can thus drop the lead index
$\nu$ and consider one effective lead: 
\begin{eqnarray}
 \H_{B} &=&  \sum_{k\sigma\alpha} \e_{k}
 c^\dagger_{k\sigma\alpha} c_{k\sigma\alpha}^{\phantom\dagger} ~.
\end{eqnarray}
Note however that the different couplings 
still enter via
the conductance prefactor $G_{0} = (e^2/h)
4\Gamma_{L}\Gamma_{R}/(\Gamma_{L}
+\Gamma_{R})^2$, which reaches the unitary limit of $e^2/h$ for a
perfect symmetric channel $\Gamma_L=\Gamma_R$
~\cite{MeirWingreen1992}, where $\Gamma_{\nu} = \pi \tilde
t^2_{\nu} \rho$ with $\rho$ the density of states of the leads at the Fermi energy.

\paragraph*{Method:}

We solve the Hamiltonian accurately using the powerful NRG
approach~\cite{Wilson75}; for a recent review see 
[\onlinecite{BullaCostiPruschke2007}].
Finite temperature NRG Green's functions are calculated using the
recently developed algorithm~\cite{PetersPruschkeAnders2006}, 
employing a complete basis set of the Wilson chain. Originally derived for
real-time dynamics of quantum impurities  out of
equilibrium~\cite{AndersSchiller2005}, this ensures 
that the spectral sum-rule is exactly fulfilled and the NRG occupancy
accurately reproduced. The self-energy of the Green function is
calculated in the usual way~\cite{BullaCostiPruschke2007}.

The local Green function determining transport is given by
$G^d_{\alpha}(z) = [z -\e -\Delta (z) -\Sigma (z)]^{-1}$ with $z=\omega + i0^{-}$,
$\Delta = \sum_{\nu} \tilde t^2_{\nu} \sum_{k}[z- \e_{k}]^{-1}$ the dot-lead 
hybridization, and $\Sigma (z)$ the interaction self-energy; 
the local spectrum is $D(\omega ,T)=\tfrac{1}{\pi}\Im m G^d_{\alpha}(z)$. 
At $T=0$, the Friedel sum rule~\cite{Langreth1966,Anders1991}
relates the number of impurity electrons $N_{imp}$ to the 
conduction electron scattering phase shift $\delta  = \pi N_{imp}/N$,
where $N$ is the number of channels (here $N=4$). $N_{imp}= N_d  +\Delta N$ 
includes all electrons on the quantum dot 
($N_d =\langle\sum_{\alpha ,\sigma}\hat{n}_{\alpha\sigma}\rangle$) \emph{and} 
the number of displaced electrons in the bath $\Delta N$~\cite{Anders1991}, 
and can  be obtained directly from the NRG~\cite{Wilson75}. For a vanishing 
$\Im m \Sigma_{\sigma\alpha}(i0^-)$ as $T\to 0$, the zero bias conductance 
must approach 
\begin{eqnarray}
  \lim_{T\to 0} G(T) &=& 4 G_{0} \sin^2(\delta) = 4 G_{0} \sin^2\left(\frac{\pi}{4} N_{imp}\right)
\label{eq:4}
\end{eqnarray}
which is indeed recovered in our calculations.

\paragraph*{Results:}

\begin{figure}
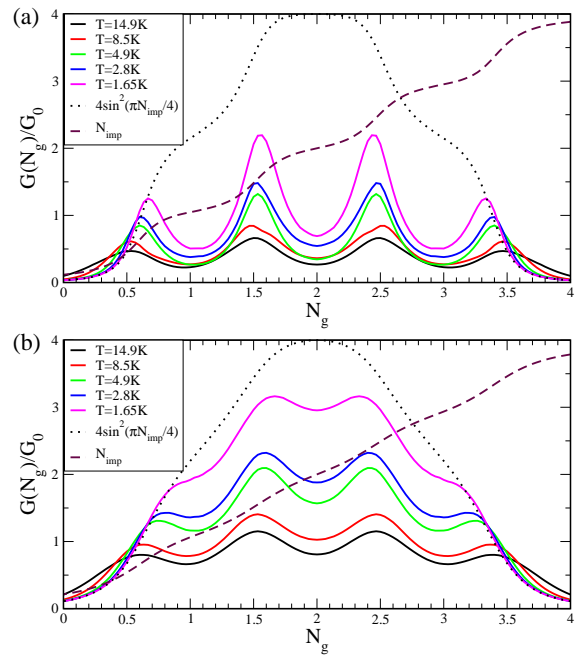

\includegraphics[width=75mm]{Fig1a.eps} \\
\includegraphics[width=75mm]{Fig1b.eps}

\caption{\label{fig:1} (color online) Zero bias conductance \emph{vs} 
    dimensionless gate voltage
    $N_g$, for different $T$
    and two tunneling strengths
    $\Gamma=0.5meV$ ((a)), $1meV$ ((b)); with $U\equiv E_C =10meV$. The
    dotted line gives the $T=0$ values,
    Eq.~(\ref{eq:4}); $N_{imp}$ is also shown (dashed line). 
    NRG parameters~\cite{Wilson75,BullaCostiPruschke2007}: $\Lambda=3$, $N_{s}=2000$.
 }
\end{figure}

We use a symmetric, constant lead density of states $\rho_0= 1/(2D)$ 
with a band width $D=30\Gamma_0$, $\Gamma_0 = 1meV \simeq 11.6K$ being
the energy unit used throughout. We take $U=10\Gamma_{0}=10meV$, corresponding 
to the experimental estimate of the charging energy $E_{C} \equiv U$~\cite{FinkelsteinII2007}.

\begin{figure}[tbh]
  \centering
 \includegraphics[width=75mm,clip]{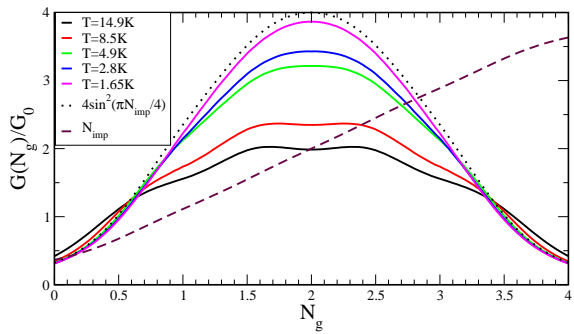}

  \caption{(color online) Zero bias conductance vs 
    $N_g$ for different temperatures and $\Gamma=2meV$.
    The dotted line gives the $T=0$ values from Eq.~(\ref{eq:4}).
    Other parameters as in Fig.~\ref{fig:1}.
  }
  \label{fig:2}
\end{figure}

One of the striking, poorly understood features of the recent
experiments on semiconducting carbon nanotube dots~\cite{FinkelsteinII2007} 
is the thermal evolution of the zero bias conductance as a function of gate
voltage.
In the experiment, the orbitals  are filled in groups of two
(there being four such groups, I-IV~\cite{FinkelsteinII2007}).
In addition, the tunneling matrix elements increase with
gate voltage. We restrict ourselves to a single representative
group, keeping the tunneling matrix elements constant for simplicity; 
and calculate the zero-bias conductance as a function of the dimensionless
gate voltage $N_g$, for different tunneling strengths $\Gamma = \Gamma_{L}+\Gamma_{R}$ 
and temperatures, using the NRG spectral functions: 
$G(T)/G_{0}=-4\int^{\infty}_{-\infty}d\omega \left(\partial f(\omega)/\partial \omega\right) 
t(\omega)$ with  $t(\omega ) = \pi \Gamma D(\omega)$ the $t$-matrix and $f(\omega)$ 
the Fermi function~\cite{MeirWingreen1992}. The results are given in Figs.~\ref{fig:1} 
and \ref{fig:2} for three different coupling strengths 
$\Gamma = 0.5\Gamma_0,\Gamma_0, 2\Gamma_0$.

Our results concur well with experiment. Those for $\Gamma=0.5meV$ (Fig.~\ref{fig:1}a) 
track the $T$-evolution of Group I depicted in Fig.~2 of Ref.~[\onlinecite{FinkelsteinII2007}].
Clear, pronounced Coulomb blockade peaks and valleys are seen in Fig.~\ref{fig:1}a.
These disappear only at temperatures much lower than the experimental base
$T=1.3K$, indicative of strong correlations ($U/\Gamma = 20$ here) and hence small 
Kondo scales $T_{K}$ in the middle of the valleys ($N_{g} = 1$ $(3)$ and $2$).
The distance between the peaks is seen to become smaller with decreasing $T$,
also in agreement with experiment. The origin is a pinning of the many-body Kondo 
resonance in $D(\omega)$ close to the chemical potential at low-$T$, evident in the t-matrix 
depicted in Fig.~\ref{fig:5} (discussed below). For $T=0$, the conductance from the Friedel sum rule Eq.~(\ref{eq:4}) is shown as a dotted line in Fig.~\ref{fig:1}a.
Absolute values of $G(T)$ also agree well with experiment.
For Group I~[\onlinecite{FinkelsteinII2007}] the coupling asymmetry is
$\Gamma_{L}/\Gamma_{R} \approx 2.5$, and hence $G_{0} \simeq 0.8e^{2}/h$.
With this, \emph{eg} the calculated $N_{g} = 1.5$ (or $2.5$) Coulomb blockade peak in $G(T)$ 
for $T=1.65K$ translates to $G_{peak} \approx 1.7e^{2}/h$, in good agreement
with experiment; and the widths of the peaks also concur with experiment.
The slight experimental asymmetry of the $N_{g} = 0.5$ and $3.5$ 
Group I conductance peaks can be attributed to a small increase in the tunnel
coupling within the Group on increasing the gate voltage.

Results for $\Gamma=1meV$ ($U/\Gamma =10$, Fig.~\ref{fig:1}b) resemble the experimental
Group III orbitals~[\onlinecite{FinkelsteinII2007}]. At higher $T$, remnants of the 
Coulomb blockade peaks remain visible; they are absent at  base temperature. While 
for stronger correlations (\emph{eg}  Fig.~\ref{fig:1}a) the charge steps are quite 
pronounced, and the different orbital filling regimes well separated in the $T=0$
conductance, for $U/\Gamma =10$ orbital filling is almost continuous.
The physical origin of such behavior is in part the `blocking effect' known from
the theory of multi-channel models in the context of rare earth materials~\cite{Bickers87}, 
in which Coulomb interactions lead to single-particle lifetime broadening of order $N\Gamma$, since $N$ relaxation channels are present. In addition, the $SU(4)$ Kondo scale in the Kondo regime is much larger than for $SU(2)$ as the number of channels enters the exponent,
$T_{K}^{SU(4)}/D \simeq [T_{K}^{SU(2)}/D]^{1/2}$~\cite{Hewson93,Galpin2005}. 
In consequence, an $SU(4)$ model enters the weakly correlated regime for 
larger $U/\Gamma$ than an $SU(2)$ symmetric  model.
\enlargethispage{\baselineskip}

Fig.~\ref{fig:2} shows results for a more weakly correlated $U/\Gamma = 5$. This
tracks the $T$-evolution of the Group IV orbitals~[\onlinecite{FinkelsteinII2007}]. Here, consistent with experiment, charge fluctuations are significant and Coulomb blockade 
peaks in consequence absent, with the $G(T\rightarrow 0)$ limit Eqn.~(\ref{eq:4}) 
being reached in practice at relatively high temperatures.

\begin{figure}[tbh]
  \centering

 \includegraphics[width=70mm]{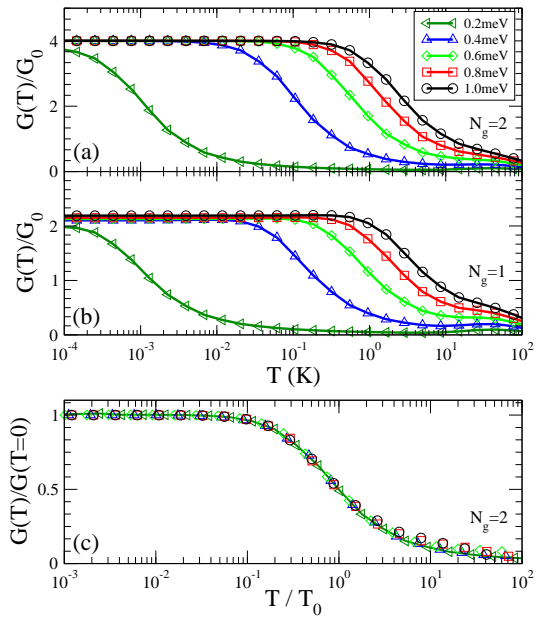}

  \caption{(color online) 
    $T$-dependence of $G(T)$ for five different 
    values of $\Gamma$ and two different 
    dimensionless gate voltages: (a) $N_{g} =2$ and (b) $N_{g}=1$ (or $N_{g}=3$).
    (c) Scaling the $N_{g} = 2$ results, $G(T)/G(0)$ \emph{vs} $x=T/T_0$.
    \label{fig:3}
  }

\end{figure}

\begin{figure}[tbh]
  \centering

  \includegraphics[width=80mm]{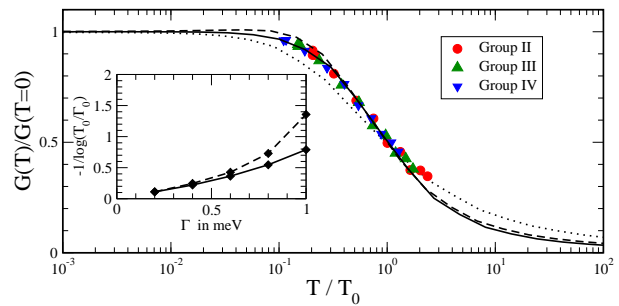}

  \caption{(color online) 
    $SU(4)$ scaling conductance
    $g(x)=G(T)/G(0)$ \emph{vs} $x=T/T_0$ 
    -- solid line for $N_g=2$, dashed for
    $N_g=1$ (or $3$). The $SU(2)$ scaling conductance is shown as a dotted line.
    Experimental results~\cite{FinkelsteinII2007} for Groups II, 
    III and IV (with $N_{g}=2$) are also shown.
    Inset: evolution of the low-energy scale $T_0$, for
    $N_{g}=2$ (solid) and $N_{g}=1$ (dashed).
  }
  \label{fig:4}
\end{figure}

\begin{figure}[thb]

 \includegraphics{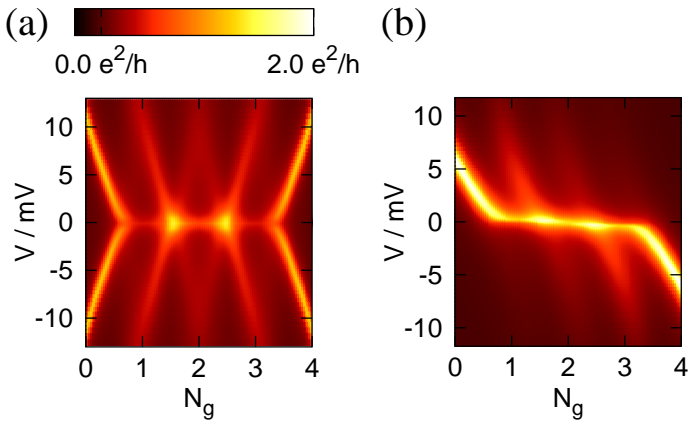}

  \caption{(color online) 
    Color-coded contour plot in the ($V,N_{g}$)-plane of (a)
    $t_{s}(V,N_{g})$ for $\Gamma = 0.5meV$ at $T=2.8K$, as in Fig.~\ref{fig:1}a;
    (b) $t(\omega = eV)$ for $\Gamma = 1meV$ as in Fig.~\ref{fig:1}b.
    \label{fig:5}
  }
\end{figure}

The $T$-dependence of $G(T)$ for the middle of the Coulomb
valleys, $N_{g}=2$ and $N_{g}=1$ (or $3$), is shown
in Fig.~\ref{fig:3}a,b for 5 values of $\Gamma$.
Comparison of the two shows the characteristic energy scales
in all valleys are of the same order (see also Fig.~\ref{fig:4} inset): 
the Coulomb blockade valleys 
are thus filled simultaneously with decreasing $T$, as seen in 
experiment~\cite{FinkelsteinII2007}. This in turn provides strong 
evidence for $U'=U$, since only a very slight decrease of $U'$ from 
the $SU(4)$ point $U'=U$ causes a sharp drop in the $N_{g}=2$ Kondo 
scale towards $SU(2)$ form~\cite{Galpin2005}, while the $N_{g}=1$ scale 
is by contrast barely affected~\cite{Galpin2005,Mitchell2006}.

Defining a characteristic low-energy scale $T_0$ $(\propto T_{K})$ by
$G(T_0)/G(0) = 0.5$, Fig.~\ref{fig:3}c shows the $N_g =2$ results from
Fig.~\ref{fig:3}a rescaled as $G(T)/G(0)$ \emph{vs} $x =T/T_{0}$. 
On progressively decreasing $\Gamma$ the data collapse, over
an ever increasing $T/T_0$-interval, onto a common curve $g(x)$.
This is the universal $SU(4)$ 
scaling conductance for $N_g =2$. That limit is well reached, for all $T/T_{0} < 10^{2}$ 
shown in Fig.~\ref{fig:3}c, by $\Gamma =0.2meV$;
but even for the largest $\Gamma$ we consider ($2meV$, $U/\Gamma =5$),  
the conductance scales onto the universal $g(x)$
over a significant $x$-range. Similar considerations arise for the $N_g =1$ results,
and the universal $SU(4)$ $g(x)$ for $N_g =2$ and $1$ (or $3$) are shown
in Fig.~\ref{fig:4}.
For $x \gtrsim 1$, $g(x)$ for $N_{g}=2$ and $1$ 
asymptotically coincide, while for lower $x$ the two are distinct (reflecting 
different leading corrections to the $SU(4)$ fixed point).
The corresponding universal $g(x)$ for a symmetric $SU(2)$ model is also shown.
It is seen to have a quite different form, and
is known~\cite{FinkelsteinII2007} not to fit well the carbon nanotube data.
For the $SU(4)$ model by contrast, our results agree
well with experiment -- as seen in Fig.~\ref{fig:4} where data for 
Groups II, III and IV in the middle of the two-electron valleys~\cite{FinkelsteinII2007}
are compared to the $N_{g}=2$ $SU(4)$ $g(x)$.
At sufficiently high $T$, departure from universal scaling must inevitably occur (as above, and just visible for the Group II, III data), 
but the data collapse remarkably well onto the scaling conductance~\footnote{Results for Group I (smallest $\Gamma$) lie slightly outside the scaling regime for most of the experimental $T$-range.}.

Finally, to assess approximately the effects of a finite source-drain bias
$V_{sd} \equiv V$, we neglect explicit dependence of the self-energy on $V$. 
Granted this, if one of the contacts is sufficiently open that it acts as a 
tunneling probe, the $T=0$ conductance $G(V,N_{g}) \propto t(\omega =eV)$ with 
$t(\omega)$ the $t$-matrix as above. This situation is appropriate to the 
Group III and IV orbitals of~[\onlinecite{FinkelsteinII2007}]. If by contrast 
the contacts are more symmetrically coupled, then
$G(V,N_{g}) \propto t_{s}(V,N_{g}) = \frac{1}{2}[t(\omega =+\frac{1}{2}eV) +
t(\omega =-\frac{1}{2}eV)]$~\cite{MeirWingreen1992}. This we believe is the 
relevant case for the Group I orbitals of~[\onlinecite{FinkelsteinII2007}].
For $\Gamma = 0.5meV$ (as in Fig.~\ref{fig:1}a), Fig.~\ref{fig:5}a shows a 
contour plot of $G_{0}t_{s}(V,N_{g})$ (calculated for $T=2.8K$~\footnote{Additional thermal smearing of $G(V,N_{g})$ from finite-$T$ Fermi functions is a minor effect.}).
The striking similarity to the experimental Group I differential conductance
map (Fig.2b of~[\onlinecite{FinkelsteinII2007}]) is evident, including the clear 
Coulomb blockade diamonds. For $\Gamma = 1meV$ (as in Fig.~\ref{fig:1}b),
Fig.~\ref{fig:5}b shows a corresponding contour plot of  $G_{0}t(\omega =eV)$; 
this in turn captures well the Group III conductance map~\cite{FinkelsteinII2007}.

\paragraph*{Conclusion:}
Motivated by recent experiments on carbon nanotube quantum dots~\cite{FinkelsteinII2007},
we have studied the evolution of an $SU(4)$ Anderson model from correlated
Kondo to mixed valent behavior, on progressive increase of the dot-lead tunnel couplings. 
NRG results obtained are in compelling agreement with experiment~\cite{FinkelsteinII2007}, from
the thermal evolution and scaling behavior of the zero-bias conductance, to differential 
conductance maps; and show that an $SU(4)$ Anderson model provides a remarkably faithful
description of carbon nanotube dots.

\begin{acknowledgments}
We acknowledge stimulating discussions with A.~Schiller. This research was 
supported in part by DFG project AN 275/5-1
and NSF Grant No. PHY05-51164 (FBA), 
by EPSRC Grant EP/D050952/1 (DEL/MRG) and by NSF Grant DMR-0239748 (GF). 

\end{acknowledgments}


\end{document}